\documentclass[12pt,a4paper]{article}

\usepackage{amssymb}
\usepackage {amsmath}
\usepackage{graphicx}
\usepackage{bm}
\usepackage{subfigure}

\begin{document}
\textwidth=135mm
 \textheight=200mm
\begin{center}
{\bfseries Two-body nucleon-nucleon correlations in Glauber-like models%
\footnote{{\small Talk presented by MR at the VI Workshop on Particle Correlations and Femtoscopy, BITP, Kiev, 14-18 September 2010.}}}
\vskip 5mm
\underline{Maciej Rybczy\'{n}ski$^{\dag}$} and Wojciech Broniowski$^{\dag,\ddag}$
\vskip 5mm
{\small {\it $^\dag$ Institute of Physics, Jan Kochanowski University, PL-25406 Kielce, Poland}} \\
{\small {\it $^\ddag$ The H. Niewodnicza\'{n}ski Institute of Nuclear Physics,
Polish Academy of Sciences, PL-31342 Krak\'{o}w, Poland}}
\\
\end{center}
\vskip 5mm
\centerline{\bf Abstract}
We investigate the role of the central two-body nucleon-nucleon correlations on 
typical quantities observed in relativistic heavy-ion collisions. Basic 
correlation measures, such as the fluctuations of the participant eccentricity, 
initial size fluctuations, or the fluctuations of the number
of sources producing particles, are sensitive at the level 10-20\% to the inclusion of 
the two-body correlations. However, the realistic correlation function gives
virtually indistinguishable results from the {\em hard-core repulsion} with the expulsion distance set to $\sim 0.9$~fm.
In the second part of the talk we compare the spherical and Gaussian wounding profiles and find that the latter, which is more 
realistic, leads to reduced eccentricity and fluctuations. This has significance for precision studies of the elliptic flow.
\vskip 10mm

Recently Refs.~\cite{alv1,alv2} published distributions
of nucleons in nuclei which account for the central Gaussian two-body nucleon-nucleon ($NN$) correlations, which is an 
important ingredient for the Glauber-model investigations \cite{glaub1,glaub2} of relativistic heavy-ion collisions. 
Up till recently the Glauber Monte Carlo codes \cite{wang,werner,bron1,alver,miller} have not
been incorporating realistic $NN$ correlations. Instead, the easy-to-implement hard-core expulsion has been used. 

The most popular model of the early stage of the collision is the {\em wounded-nucleon model} \cite{bialas1}.
Variants of the approach \cite{bron1,kharzeev1,back1,back2} admix a certain fraction of binary collisions
to the wounded nucleons, which leads to a better overall description of multiplicities of the produced 
particles. In this {\em mixed model}, used throughout the talk,
the number of the produced particles is proportional to the number of {\em sources},
\begin{eqnarray}
N_s=(1-\alpha)N_{w}/2+\alpha N_{\rm bin},
\end{eqnarray}
where $N_w$ is the number of the wounded nucleons (those who interacted inelastically at least once) 
and $N_{\rm bin}$ denotes the number of binary collisions.
More sophisticated approaches \cite{bozek1} discriminate between the nucleons which have collided only once (corona) and more than once (core). 
Also, the wounded-quark model \cite{bialas2,bialas3,bzdak1} yields a quite successful phenomenology.
In this talk we apply the mixed model for the $^{208}$Pb-$^{208}$Pb collisions with $\alpha=0.12$, corresponding to the highest SPS energy. 
\vskip 10mm

\begin{figure}
\begin{center}
\includegraphics[width=0.5\textwidth]{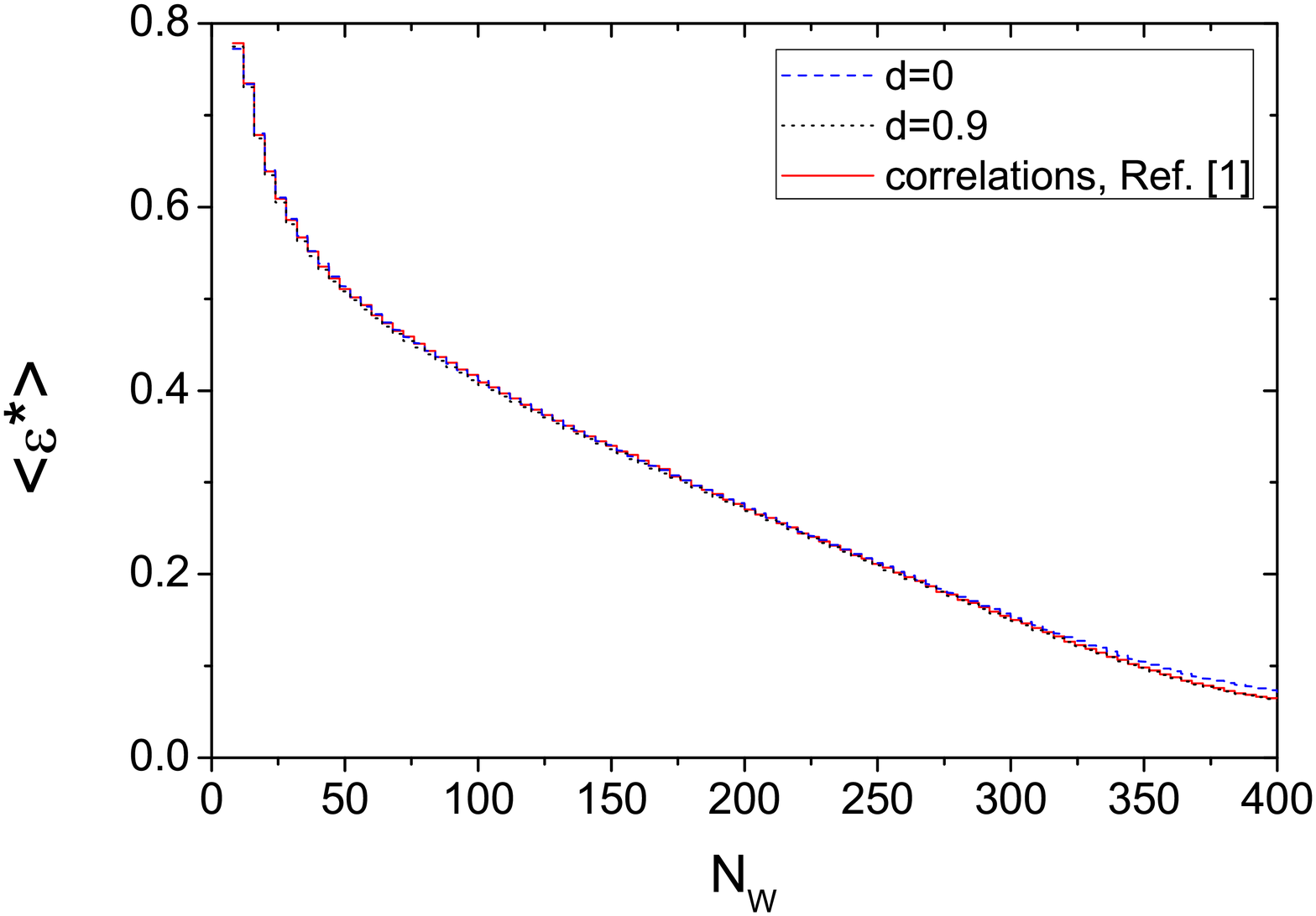}
\includegraphics[width=0.5\textwidth]{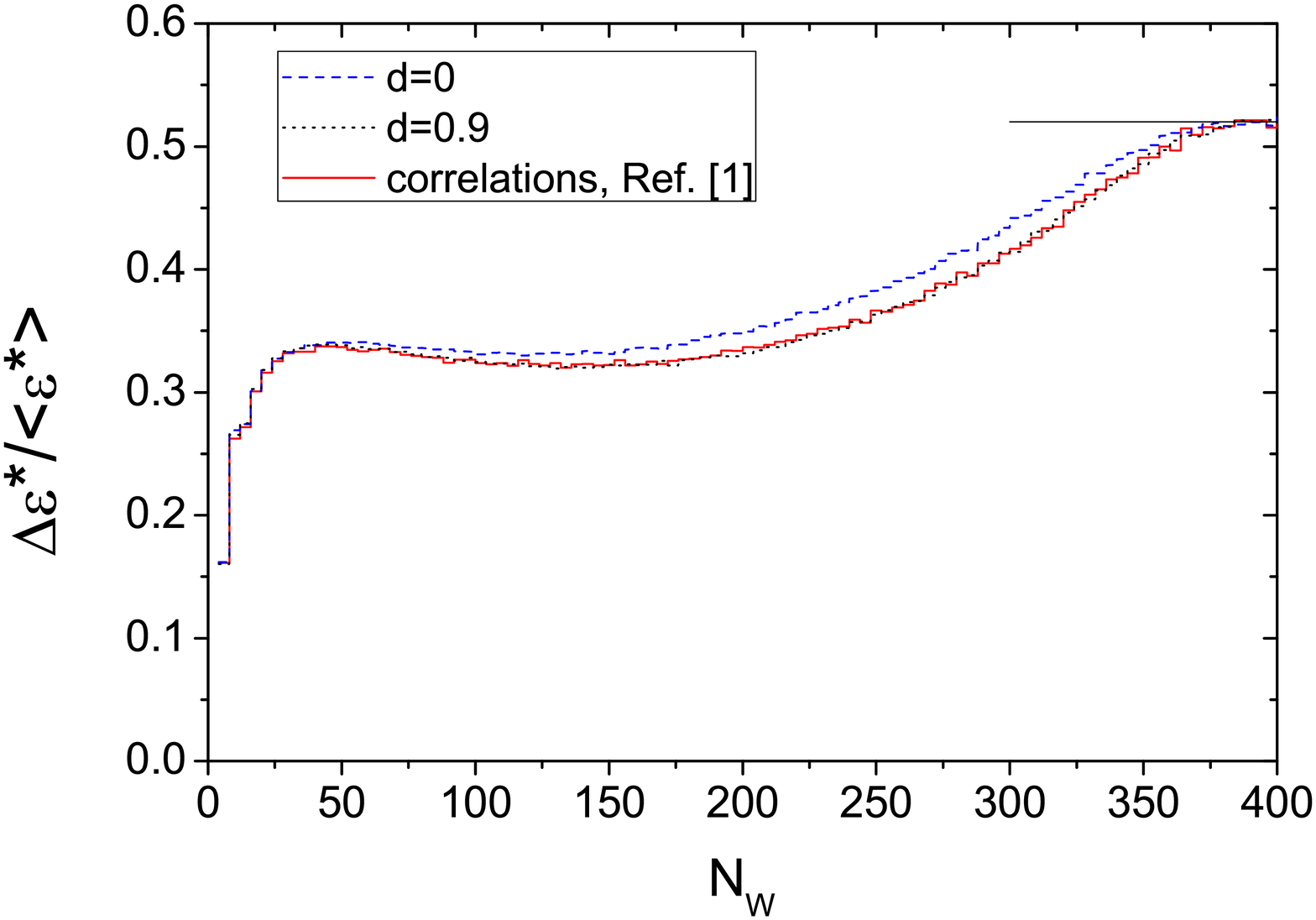}
\end{center}
\vspace{-6mm}
\caption{Left: the average participant eccentricity, $\langle \varepsilon^\ast \rangle$, plotted 
vs the number of wounded nucleons, $N_w$, obtained with the three investigated nucleon distributions described in the text.
Right: The scaled standard deviation, $\Delta \varepsilon^\ast/\langle \varepsilon^\ast \rangle$, obtained from an event-by-event
study. The short horizontal line at the most central events corresponds to the value $\sqrt{4/\pi-1}$
of Ref.~\cite{bron3} following from the central limit theorem. $^{208}{\rm Pb}-{}^{208}{\rm Pb}$ collisions. 
\label{fig:eps}}
\end{figure}

First, we compare the results of the Glauber calculation initialized with the correlated distributions of Ref.~\cite{alv1,alv2} 
(solid lines in the figures), with uncorrelated distributions (dashed lines), and with the distributions accounting for the hard-core repulsion
with the expulsion radius $d=0.9$~fm (dotted lines).
The Monte Carlo simulations are performed with {\tt GLISSANDO} \cite{bron1}. 
We start with the  {\em participant} eccentricity, $\varepsilon^\ast$, appearing in the studies of the event-by-event 
fluctuations of the initial shape, 
in particular of its elliptic component \cite{bhal,andrade}. 
In the left panel of Fig.~\ref{fig:eps} we show the dependence of the event-by-event average,
$\langle \varepsilon^\ast \rangle$, on $N_w$. We note that the three calculations are virtually
indistinguishable, except for a tiny difference for the most central collisions, where the uncorrelated case is a few percent higher. The same conclusions were reached in the analogous study of eccentricity in Ref.~\cite{tavares}.

The right panel of Fig.~\ref{fig:eps} shows the scaled standard deviation of the participant eccentricity,
$\Delta \varepsilon^\ast/\langle \varepsilon^\ast \rangle$, obtained from our event-by-event
analysis. We note a significant difference between the
uncorrelated case, which has up to 10\% larger fluctuations at intermediate
centralities, and the cases with correlations. 
However, the calculations with the realistic $NN$ correlations and the hard-core
correlations give a virtually indistinguishable result, with the two curves overlapping within the
statistical noise. The short horizontal line at the most central events corresponds to the value $\sqrt{4/\pi-1}$
of Ref.~\cite{bron3}, following from the central limit theorem for the most central events.
\vskip 10mm

Now we pass to the second part of the talk.
In the existing Glauber Monte-Carlo codes there is a common use of the spherical wounding profiles. In other words, the collision 
occurs when the transverse distance between the centers of the colliding nucleons, $b$, is less then $R$, where, 
geometrically, $\pi R^{2}=\sigma_{inel}$. We can also write that the collision probability distribution in $b$ (the wounding 
profile) has the 
form $\sigma(b)=\Theta(R-b)$.
However, it was shown in Ref.~\cite{bialas0} that the use of the wounding profile in the Gaussian form,
\begin{eqnarray}
\sigma(b)=A\exp\left(\frac{-Ab^{2}}{R^{2}} \right),
\end{eqnarray}
with $A=0.92$ tuned to the $NN$ scattering data,
leads to much more realistic results. In particular, a combination of gaussians can explain in detail the nucleon-nucleon elastic 
cross section, including its diffractive features. Although the integrated $NN$ cross section is by construction the same for the 
Gaussian and the hard-sphere wounding profiles, $\int d^2 b \,\sigma(b)=\pi R^2$, the Gaussian profile has a tail, making a collision of 
distant nucleons possible.

We have investigated the influence of the Gaussian wounding profile on the participant eccentricity 
and on the scaled variance of the number of sources, $\omega_{S}$, defined as 
\begin{eqnarray}
\omega_{s}=\frac{{\rm Var}(N_{S})}{\langle N_{S}\rangle}.
\end{eqnarray}

\begin{figure}
\begin{center}
\includegraphics[width=0.5\textwidth]{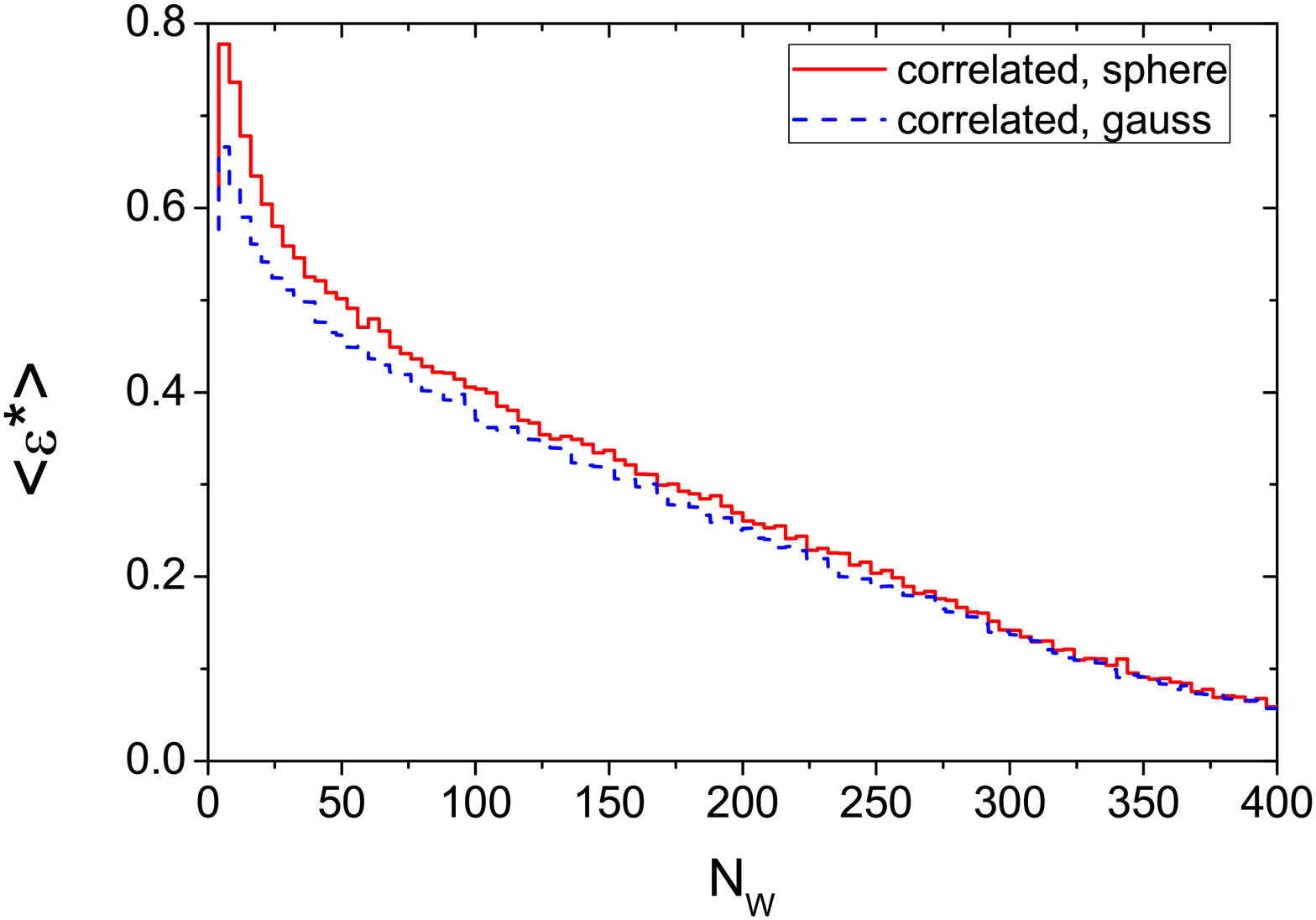}
\includegraphics[width=0.5\textwidth]{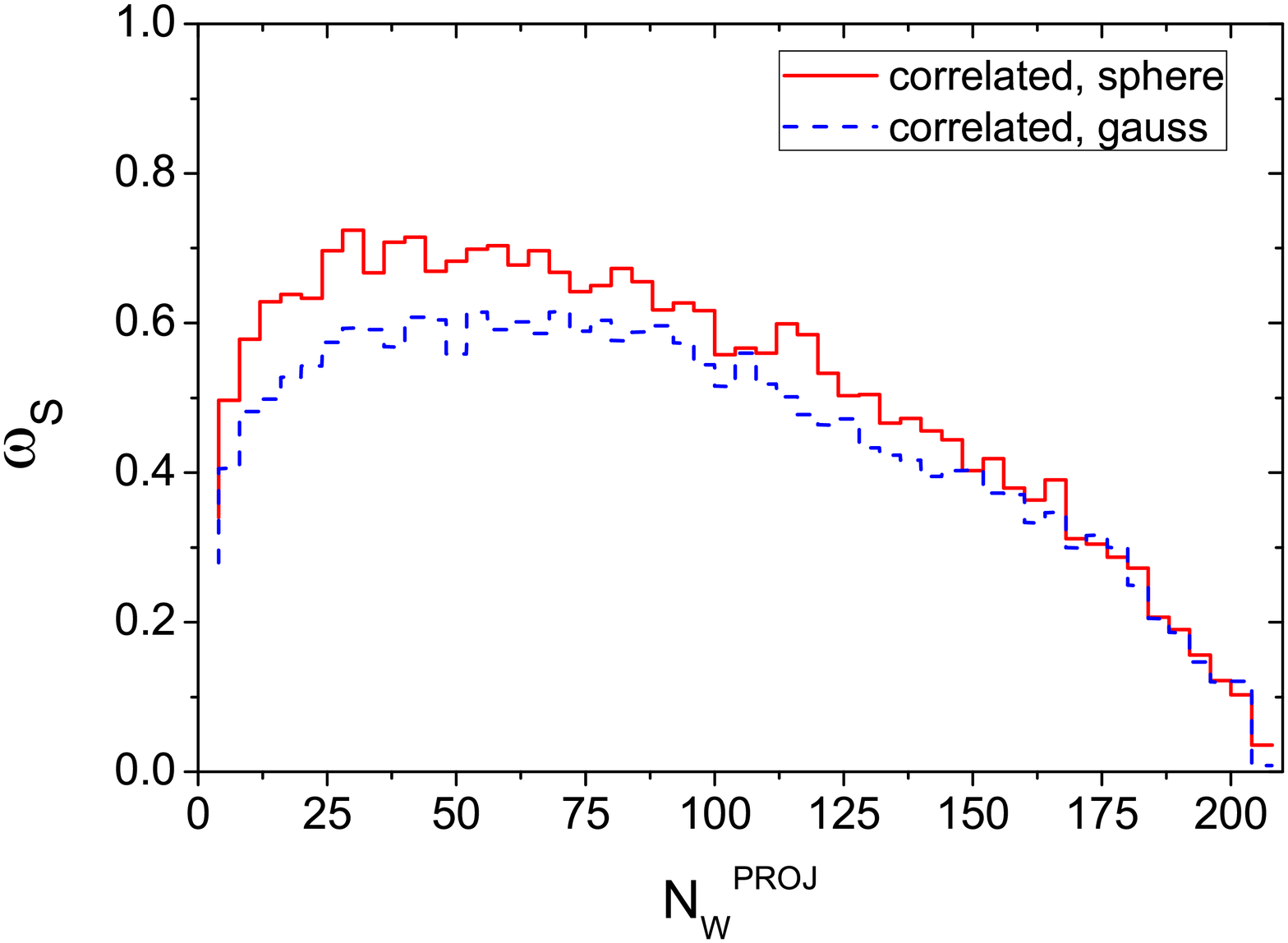}
\end{center}
\vspace{-6mm}
\caption{Left: the average participant eccentricity, $\langle \varepsilon^\ast \rangle$, plotted 
vs the number of wounded nucleons, $N_w$, obtained with the correlated \cite{alv1} nucleon distribution 
for the spherical and Gaussian wounding profiles (see the text for details).
Right: The scaled variance of the number of sources, $\omega_{S}$,
plotted as a function of the number of wounded nucleons in the projectile, $N_W^{\rm PROJ}$, for the case 
two wounding profiles.
\label{fig:gaussian}}
\end{figure}

In the left panel of the Fig.~\ref{fig:gaussian} we have plotted the average participant eccentricity, $\langle \varepsilon^\ast \rangle$, 
vs the number of wounded nucleons, $N_w$, obtained with the correlated \cite{alv1} nucleon distribution for the spherical and Gaussian
wounding profiles. There is a visible $10-15\%$ quenching of $\langle \varepsilon^\ast \rangle$ seen in peripheral collisions 
for the Gaussian wounding profile in comparison to the spherical profile. The simple explanation of the fact 
is that with more extended wounding profile the in-plane nucleons have a larger chance to become wounded, which decreases 
$\langle \varepsilon^\ast \rangle$. This quenching, although not very large, has significance for precision studies of 
the elliptic-flow coefficient, $v_2$, which in hydrodynamic studies is sensitive to the initial eccentricity. 
While taking into account the fluctuations (participant eccentricity) increases $\langle \varepsilon^\ast \rangle$, the 
use of the realistic wounding profile brings it down.
 
The right panel of the Fig.~\ref{fig:gaussian} shows the scaled variance of the number of sources, 
$\omega_{S}$, plotted as a function of the number of wounded nucleons in the projectile, $N_W^{\rm PROJ}$. Here we also note a 
decrease of the fluctuations when the Gaussian wounding profile is used.
\vskip 10mm

The current version of the {\tt GLISSANDO} package can be downloaded from {\tt http://www.ujk.edu.pl/homepages/mryb/GLISSANDO/index.html}.

\vskip 15mm

Research supported by Polish Ministry of Science and Higher Education, grants N~N202~263438 and N~N202~249235.

\end{document}